\documentclass[journal,10pt]{IEEEtran}
\usepackage[T1]{fontenc}
\usepackage{cite}
\usepackage{amsmath,amssymb}
\usepackage{graphicx}
\usepackage{booktabs}
\usepackage{array}
\usepackage{multirow}
\usepackage{siunitx}
\usepackage{xcolor}
\usepackage{url}
\usepackage{microtype}
\usepackage{balance}
\usepackage{placeins}
\usepackage{float}
\usepackage{tikz}
\usetikzlibrary{arrows.meta,positioning,fit,calc}

\graphicspath{{figures/}}
\newcommand{\CN}{\mathcal{CN}}
\newcommand{\diag}{\operatorname{diag}}
\newcommand{\argmax}{\operatorname*{arg\,max}}

\newcommand{\vect}[1]{\boldsymbol{#1}}
\newcommand{\mat}[1]{\mathbf{#1}}

\newcommand{\indicator}{\mathbb{I}}

\newcommand{\QuantizationMeanErr}{0.380}

\newcommand{\CentroidNativeAcc}{65.20}
\newcommand{\CentroidNativeErr}{0.735}
\newcommand{\CentroidZPTwoAcc}{78.20}
\newcommand{\CentroidZPTwoErr}{0.496}
\newcommand{\CentroidZPThreeAcc}{79.43}
\newcommand{\CentroidZPThreeErr}{0.477}
\newcommand{\CentroidSBLTwoAcc}{84.29}
\newcommand{\CentroidSBLTwoErr}{0.440}
\newcommand{\CentroidSBLThreeAcc}{85.06}
\newcommand{\CentroidSBLThreeErr}{0.421}

\newcommand{\KNNNativeAcc}{75.79}
\newcommand{\KNNNativeErr}{0.587}
\newcommand{\KNNZPTwoAcc}{86.12}
\newcommand{\KNNZPTwoErr}{0.431}
\newcommand{\KNNZPThreeAcc}{87.24}
\newcommand{\KNNZPThreeErr}{0.419}
\newcommand{\KNNSBLTwoAcc}{93.27}
\newcommand{\KNNSBLTwoErr}{0.392}
\newcommand{\KNNSBLThreeAcc}{93.13}
\newcommand{\KNNSBLThreeErr}{0.394}

\newcommand{\LinearNativeAcc}{76.36}
\newcommand{\LinearNativeErr}{0.542}
\newcommand{\LinearZPTwoAcc}{85.42}
\newcommand{\LinearZPTwoErr}{0.421}
\newcommand{\LinearZPThreeAcc}{86.43}
\newcommand{\LinearZPThreeErr}{0.414}
\newcommand{\LinearSBLTwoAcc}{88.60}
\newcommand{\LinearSBLTwoErr}{0.408}
\newcommand{\LinearSBLThreeAcc}{89.18}
\newcommand{\LinearSBLThreeErr}{0.401}

\newcommand{\RBFNativeAcc}{77.39}
\newcommand{\RBFNativeErr}{0.517}
\newcommand{\RBFZPTwoAcc}{85.30}
\newcommand{\RBFZPTwoErr}{0.421}
\newcommand{\RBFZPThreeAcc}{86.15}
\newcommand{\RBFZPThreeErr}{0.415}
\newcommand{\RBFSBLTwoAcc}{90.18}
\newcommand{\RBFSBLTwoErr}{0.398}
\newcommand{\RBFSBLThreeAcc}{90.23}
\newcommand{\RBFSBLThreeErr}{0.398}

\newcommand{\RFNativeAcc}{81.38}
\newcommand{\RFNativeErr}{0.464}
\newcommand{\RFZPTwoAcc}{87.07}
\newcommand{\RFZPTwoErr}{0.422}
\newcommand{\RFZPThreeAcc}{88.05}
\newcommand{\RFZPThreeErr}{0.410}
\newcommand{\RFSBLTwoAcc}{89.29}
\newcommand{\RFSBLTwoErr}{0.406}
\newcommand{\RFSBLThreeAcc}{89.92}
\newcommand{\RFSBLThreeErr}{0.401}

\begin{document}
\title{Single-Base-Station Indoor Localization via Super-Resolved Relative Power Delay Profiles}

\author{Fangqing~Xiao,~\IEEEmembership{Member,~IEEE,}
        and~Dirk~T.~M.~Slock,~\IEEEmembership{Life~Fellow,~IEEE}%
\thanks{F. Xiao is with the School of Information Science and Engineering, Yunnan University, Kunming 650091, China (e-mail: fangqing.xiao@ynu.edu.cn).}%
\thanks{D. T. M. Slock is with the Communication Systems Department, EURECOM, 06410 Biot, France (e-mail: dirk.slock@eurecom.fr).}%
\thanks{Corresponding author: Fangqing Xiao.}}

\markboth{IEEE Wireless Communications Letters,~Vol.~XX, No.~XX, 2026}%
{Xiao and Slock: Single-BS Localization via Super-Resolved Relative PDPs}

\maketitle

\begin{abstract}
Indoor multipath is shaped by surrounding reflectors, scatterers, and blockages, so a relative power-delay profile (PDP) can serve as a location fingerprint without an identifiable LoS path, angle information, or absolute time-of-arrival ranging. However, a communication receiver observes finitely many noisy pilot-frequency samples rather than an ideal PDP. This paper models the resulting Dirichlet blur, delay folding, and off-grid mismatch, and reconstructs a posterior-power profile using expectation-maximization sparse Bayesian learning. In spatially consistent QuaDRiGa simulations, twofold SBL raises 20-dB Top-1 accuracy from 75.79\% (native PDP) and 87.24\% (threefold zero-padding) to 93.27\%, with 0.392~m mean error.
\end{abstract}

\begin{IEEEkeywords}
Indoor localization, OFDM channel sounding, power delay profile, radio-map localization, single base station.
\end{IEEEkeywords}

\section{Introduction}
\IEEEPARstart{R}{adio}-map localization matches an online channel feature to labeled measurements collected offline. Classical systems rely mainly on received signal strength from multiple access points \cite{bahl2000radar,he2016wifi}, whereas accurate single-access-point systems commonly exploit antenna-array angles or calibrated time of flight \cite{kotaru2015spotfi,vasisht2016chronos}. A power delay profile (PDP) offers a different route. In an indoor environment, the relative delays and powers of multipath components are shaped by the surrounding geometry, reflectors, scatterers, and blockage conditions. Hence, even when no LoS path can be reliably identified and no absolute propagation time is used, the relative PDP shape can remain highly location-dependent. The key idea is to use NLOS multipath as environmental information for localization rather than treating it only as an impairment. A line of work initiated single-link PDP fingerprinting and subsequently studied its NLOS interpretation, delay--Doppler extension, and indoor validation \cite{triki2006pdp,triki2007nlos,oktem2010pddp,ding2016pdp}. Unlike model-based PDP ranging or range--Doppler sensing, which exploit a propagation or sensing model and may benefit from a LoS-related reference \cite{xiao2024ranging,xiao2026pdpsensing}, the present work uses the PDP only as a relative radio-map fingerprint.

A remaining gap is that a communication receiver does not observe an ideal continuous-delay PDP. It observes finitely many noisy channel samples on a pilot-frequency grid. Commodity-WiFi work has shown that bandwidth and hardware effects can strongly limit a directly transformed PDP \cite{xie2019pdp}, while sparse channel-estimation methods recover multipath structure from limited training data \cite{bajwa2010ccs,qiao2018sbl}. More fundamentally, bandlimited Fourier observations admit super-resolution only through structural assumptions, and a discretized delay dictionary introduces off-grid mismatch \cite{candes2014superresolution,tang2013offgrid,yang2013offgrid}. These channel-estimation studies optimize reconstruction fidelity, not the discriminability of a radio-map feature. Conversely, prior PDP-localization studies generally begin after a PDP has already been formed and do not isolate how finite aperture, uniform pilot spacing, and delay-grid choice reshape the fingerprint. Recent super-resolution CSI fingerprinting uses angle--delay--power features from massive-MIMO measurements \cite{uykan2026superresolution}; the present problem is SISO, contains no angle information, does not estimate range, does not associate any recovered peak with a physical LoS path, and constructs a relative PDP directly from the same finite pilot observations available to every compared front end.

This paper addresses that observation-to-localization gap. First, it formulates LoS-free single-BS localization from pilot-domain relative PDPs, treating multipath as a source of environmental information rather than as an impairment. Second, it derives the periodic Dirichlet point-spread function induced by uniformly spaced pilots and explicitly retains the pilot-grid-induced folded-delay representation. Folding loses the identity of unfolded physical delays, but it does not remove their contribution to the measured fingerprint; the same folded representation is learned offline and used online. Third, it forms the fingerprint from the EM-SBL posterior second moment and compares it with the native PDP and dimension-matched twofold and threefold zero-padding baselines. Spatially consistent full-channel simulations then evaluate both reconstruction grids through SNR sweeps, error CDFs at 10 and 20~dB, and five conventional localizers, reporting both cell accuracy and metric position error.

\section{Pilot-Domain Relative-PDP Observation Model}
\begin{figure}[htpb]
	\centering
	\includegraphics[width=0.6\columnwidth]{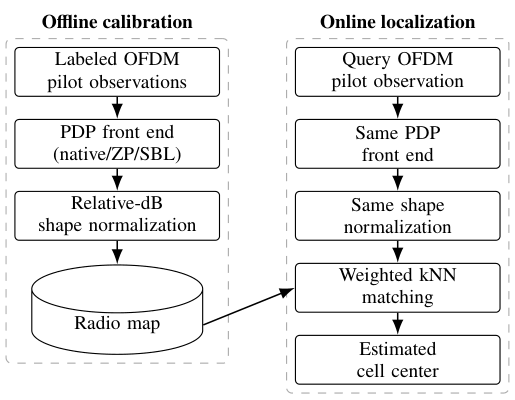}
	\caption{Offline radio-map construction and online single-BS localization. The two stages use the same pilot processing, PDP front end, feature support, and shape normalization; only the online stage performs radio-map matching.}
	\label{fig:workflow}
\end{figure}

Fig.~\ref{fig:workflow} separates the proposed localization chain into a pilot-domain PDP front end and a conventional radio-map inference stage. This section establishes what can be inferred before any particular reconstruction algorithm is chosen. We first define the relative-delay observation, then characterize the two deterministic consequences of finite uniformly spaced pilots: periodic delay folding and finite-aperture blur.
\subsection{Relative-delay pilot observation}
Consider a fixed SISO base station (BS) and a user equipment (UE) at horizontal position $\vect{p}\in\mathbb{R}^{2}$. The physical channel is
\begin{equation}
 h^{\rm abs}_{\vect{p}}(\tau)=\sum_{\ell=1}^{L_{\vect{p}}}
 \alpha_{\vect{p},\ell}\delta(\tau-\tau^{\rm abs}_{\vect{p},\ell}).
 \label{eq:absolute_channel}
\end{equation}
Conventional packet detection and coarse OFDM timing supply a reproducible receiver-side delay origin \cite{schmidl1997sync,minn2003sync}. Denoting it by $\tau_{\vect{p},0}$, the aligned delays are
\begin{equation}
 \tau_{\vect{p},\ell}=\tau^{\rm abs}_{\vect{p},\ell}-\tau_{\vect{p},0}.
 \label{eq:relative_delay}
\end{equation}
The origin need not be associated with a prescribed LoS, strongest, or individually identified physical path. It only has to be generated by the same communication-alignment convention during offline calibration and online operation. Localization therefore uses the relative multipath shape, not the absolute propagation time or an identified LoS component.

Let $M$ unit-modulus sounding pilots occupy the frequency offsets
\begin{equation}
 \nu_m=\nu_0+m\Delta f_{\rm p},\qquad m=0,\ldots,M-1.
 \label{eq:pilot_grid}
\end{equation}
After division by the known pilots, the channel samples are
\begin{equation}
 y_m=\frac{1}{\sqrt M}\sum_{\ell=1}^{L_{\vect{p}}}\alpha_{\vect{p},\ell}
 e^{-\mathrm{j}2\pi \nu_m\tau_{\vect{p},\ell}}+w_m,
 \quad w_m\sim\CN(0,\sigma_w^2).
 \label{eq:pilot_observation}
\end{equation}
All PDP front ends operate on this same vector $\vect{y}\in\mathbb{C}^{M}$. The sounding interval is assumed quasi-static and protected by a sufficient guard interval. The pilot spacing and the finite number of observed tones now determine, respectively, the identifiable delay interval and the within-interval resolution.

\subsection{Delay folding induced by the pilot spacing}
Uniform spacing makes \eqref{eq:pilot_observation} periodic in delay with ambiguity period
\begin{equation}
 T_{\rm amb}=\frac{1}{\Delta f_{\rm p}}.
 \label{eq:ambiguity_period}
\end{equation}
Writing $\tau_{\vect{p},\ell}=\bar\tau_{\vect{p},\ell}+k_{\vect{p},\ell}T_{\rm amb}$, where $\bar\tau_{\vect{p},\ell}\in[0,T_{\rm amb})$ and $k_{\vect{p},\ell}\in\mathbb{Z}$, gives
\begin{equation}
 e^{-\mathrm{j}2\pi\nu_m\tau_{\vect{p},\ell}}
 =e^{-\mathrm{j}2\pi\nu_0k_{\vect{p},\ell}T_{\rm amb}}
 e^{-\mathrm{j}2\pi\nu_m\bar\tau_{\vect{p},\ell}}.
 \label{eq:alias_equivalence}
\end{equation}
Thus paths separated by integer multiples of $T_{\rm amb}$ are not individually identifiable from this pilot grid; the remaining constant phase is absorbed into the complex gain. We do not claim physical delay unaliasing. Instead, the folded relative PDP over one ambiguity period is used as a position-dependent fingerprint. Because offline and online observations use the same pilot grid and alignment convention, this folding mechanism is identical in both stages. Equation~\eqref{eq:ambiguity_period} fixes the recoverable delay domain; the finite aperture determines how sharply paths can be represented inside that domain.

\subsection{Dirichlet blur induced by the finite aperture}
Define the continuous matched-filter response
\begin{equation}
 r(\tau)=\frac{1}{\sqrt M}\sum_{m=0}^{M-1}y_m e^{\mathrm{j}2\pi\nu_m\tau}.
 \label{eq:matched_filter}
\end{equation}
Define $\widetilde r(\tau)=e^{-\mathrm{j}2\pi\nu_0\tau}r(\tau)$, for which $|\widetilde r(\tau)|^2=|r(\tau)|^2$. Substitution of \eqref{eq:pilot_observation} gives
\begin{equation}
 \widetilde r(\tau)=\sum_{\ell}\beta_{\vect{p},\ell}
 D_M\!\left(\Delta f_{\rm p}(\tau-\bar\tau_{\vect{p},\ell})\right)+\widetilde w(\tau),
 \label{eq:dirichlet_distortion}
\end{equation}
where $\beta_{\vect{p},\ell}=\alpha_{\vect{p},\ell}e^{-\mathrm{j}2\pi\nu_0\tau_{\vect{p},\ell}}$ absorbs the reference-frequency phase and $\widetilde w(\tau)=M^{-1/2}e^{-\mathrm{j}2\pi\nu_0\tau}\sum_{m=0}^{M-1}w_m e^{\mathrm{j}2\pi\nu_m\tau}$. The periodic kernel is
\begin{equation}
 D_M(u)=\frac{1}{M}\sum_{m=0}^{M-1}e^{\mathrm{j}2\pi m u}
 =e^{\mathrm{j}\pi(M-1)u}\frac{\sin(\pi M u)}{M\sin(\pi u)}.
 \label{eq:dirichlet_kernel}
\end{equation}
Finite aperture therefore broadens each component, generates sidelobes, and causes nearby paths to combine coherently. Evaluating $r(\tau)$ only on the native inverse-DFT grid introduces an additional delay-bin quantization. Zero-padding merely samples the same blurred response more densely; it does not alter $D_M$, increase the observed bandwidth, or add measurements. These distortions motivate a sparse posterior-power representation constructed from the original pilot samples.

\section{Sampling-Aware PDP Fingerprinting and Radio-Map Inference}
Based on the observation model, we define three PDP front ends and a common feature/matching protocol so that performance differences isolate the representation rather than the downstream localizer.
\subsection{Native, zero-padded, and EM-SBL PDPs}
Define $[\vect{a}(\tau)]_m=M^{-1/2}e^{-\mathrm{j}2\pi\nu_m\tau}$. The native delay-bin spacing is
\begin{equation}
 \Delta\tau_{\rm N}=\frac{1}{M\Delta f_{\rm p}}.
 \label{eq:native_bin}
\end{equation}
The native PDP is $\widehat p_{\rm N}[n]=|r(n\Delta\tau_{\rm N})|^2$, sampled on the natural $M$-point delay grid. For $q\in\{2,3\}$, $q$-fold zero-padding samples the same dirty response on a $qM$-point grid from the same $M$ pilots; it adds neither observations nor resolving power and serves as the dimension-matched control for $q$-fold SBL.

For an oversampling factor $q\geq 1$, define $N_q=qM$, $\tau_n^{(q)}=n\Delta\tau_{\rm N}/q$ for $n=0,\ldots,N_q-1$, and
\begin{equation}
 \mat{A}_q=\big[\vect{a}(\tau_0^{(q)}),\ldots,
 \vect{a}(\tau_{N_q-1}^{(q)})\big].
 \label{eq:sbl_dictionary}
\end{equation}
Here ``super-resolved'' denotes sparse posterior-power reconstruction on a grid finer than $\Delta\tau_{\rm N}$; it does not imply physical-delay unaliasing beyond $T_{\rm amb}$. Let $s_y=\|\vect{y}\|_2^2/M$, $\bar{\vect{y}}=\vect{y}/\sqrt{s_y}$, and $\bar\sigma_w^2=\underline\sigma_w^2/s_y$, where $\underline\sigma_w^2$ is the calibrated thermal-noise floor before normalization. EM-SBL is applied to $\bar{\vect{y}}$; to lighten notation, the normalized observation is written as $\vect{y}$ below. Using the automatic-relevance-determination model of SBL \cite{tipping2001sbl,wipf2004sbl},
\begin{equation}
 \vect{y}=\mat{A}_q\vect{x}_q+\vect{w},\qquad
 \vect{w}\sim\CN(\vect{0},\sigma^2\mat{I}),
\end{equation}
with
\begin{equation}
 p(\vect{x}_q\mid\vect{\gamma})=\CN(\vect{0},\mat{\Gamma}),
 \quad \mat{\Gamma}=\diag(\vect{\gamma}).
 \label{eq:sbl_model}
\end{equation}
With $\mat{C}=\mat{A}_q\mat{\Gamma}\mat{A}_q^{\mathsf H}+\sigma^2\mat{I}$, the posterior moments are
\begin{align}
 \vect{\mu}=\mat{\Gamma}\mat{A}_q^{\mathsf H}\mat{C}^{-1}\vect{y}, \ \ 
 \mat{\Sigma}=\mat{\Gamma}-\mat{\Gamma}\mat{A}_q^{\mathsf H}
 \mat{C}^{-1}\mat{A}_q\mat{\Gamma}.
 \label{eq:sbl_posterior}
\end{align}
The EM sufficient statistics are
\begin{align}
 \widetilde\gamma_n=|\mu_n|^2+[\mat{\Sigma}]_{n,n},\ \ 
 \widetilde\sigma^2=\frac{\|\vect{y}-\mat{A}_q\vect{\mu}\|_2^2+
 \operatorname{tr}(\mat{A}_q\mat{\Sigma}\mat{A}_q^{\mathsf H})}{M}.
 \label{eq:sbl_em_update}
\end{align}
Damped EM updates both the relevance parameters and the unknown effective residual variance:
\begin{align}
 \gamma_n^{(t+1)}&=(1-\rho)\gamma_n^{(t)}+\rho\widetilde\gamma_n^{(t)},\nonumber\\
 \sigma_{t+1}^2&=\Pi_{[\sigma_{\min}^2,\sigma_{\max}^2]}
 \!\left((1-\rho)\sigma_t^2+\rho\widetilde\sigma_t^2\right).
 \label{eq:damped_updates}
\end{align}
Here $\Pi$ denotes interval projection. For the normalized data, set $\sigma_{\min}^2=\max\{\bar\sigma_w^2,10^{-12}\}$ and $\sigma_{\max}^2=\max\{10^{-1},10\sigma_{\min}^2\}$. The learned excess above the normalized thermal-noise floor absorbs part of the continuous-delay/grid mismatch. Let $g_n=|\vect{a}_n^{\mathsf H}\vect{y}|^2$ and $\bar g_n=Mg_n/\sum_k g_k$. We initialize $\gamma_n^{(0)}=\max\{\bar g_n,10^{-10}\max_j\bar g_j\}$ and $\sigma_0^2=\Pi_{[\sigma_{\min}^2,\sigma_{\max}^2]}(10^{-4})$. Delayed pruning only accelerates the iterations. A measurement-space Cholesky implementation costs $\mathcal O(M^3+M^2N_q)$ per iteration. The reported SBL PDP is explicitly the posterior second moment
\begin{equation}
 \widehat p_{\rm SBL}[n]=|\mu_n|^2+[\mat{\Sigma}]_{n,n},
 \label{eq:sbl_pdp}
\end{equation}
which equals the relevance profile at an EM fixed point. Since an off-grid physical path can spread over adjacent atoms, \eqref{eq:sbl_pdp} is a refined power representation rather than a one-to-one physical-path estimate \cite{tang2013offgrid,yang2013offgrid}.

Unlike zero-padding, EM-SBL infers a sparse latent posterior from the pilots; matched-$q$ comparisons therefore separate sparse inference from interpolation and feature dimension.

\subsection{Feature construction and offline/online inference}
Every front end retains its complete sampled representation over the ambiguity interval $[0,T_{\rm amb})$; no delay-window truncation is applied. For the nonzero reconstructed profile on delay grid $\{\tau_n\}_{n=0}^{N-1}$, define $p_{\max}=\max_{0\leq k<N}\widehat p[k]$ and the common post-reconstruction clipping level $p_{\min}=p_{\max}10^{-B_{\rm dB}/10}$. The shape-only feature is
\begin{equation}
 z_n=1+\frac{10}{B_{\rm dB}}\log_{10}
 \frac{\max\{\widehat p[n],p_{\min}\}}{p_{\max}},
 \qquad 0\leq n<N.
 \label{eq:normalization}
\end{equation}
Thus $z_n\in[0,1]$, absolute received power is removed, and reconstructed powers more than $B_{\rm dB}$ below the profile maximum are mapped to zero. All delay bins remain present in the feature vector; this clipping does not truncate the delay interval or remove physical paths before pilot sampling. Each representation uses its own consistent training/validation/test front end and separate radio map (and classifier in Table~\ref{tab:classifiers}); data splits, channels, paired noise draws, and selection rules are shared.

The service area is divided into $C$ cells with centers $\{\vect{c}_c\}_{c=1}^{C}$, and the radio map contains labeled pairs $(\vect{z}_i,c_i)$. For a query $\vect{z}$, let $\mathcal N_K(\vect{z})$ contain the $K$ nearest features in Euclidean distance. The inverse-distance class score and reported position are
\begin{equation}
 s_c(\vect{z})=\sum_{i\in\mathcal N_K(\vect{z})}
 \frac{\indicator\{c_i=c\}}{\|\vect{z}-\vect{z}_i\|_2+10^{-6}},
 \quad
 \widehat{\vect{p}}=\vect{c}_{\argmax_c s_c(\vect{z})}.
 \label{eq:weighted_knn}
\end{equation}
For each representation, $K$ is selected independently from $\{1,3,5,7,9,11,15\}$ using two 20-dB validation realizations per coordinate and then frozen for all test SNRs, isolating the PDP front end from the decision rule \cite{he2016wifi}.

\section{Simulation Results}
The evaluation follows the processing order in Fig.~\ref{fig:workflow}: we first instantiate the map, pilot grid, full ambiguity-period feature support, and EM settings; next we inspect the native, zero-padded, and SBL representations of one channel; finally we compare the principal SNR curves, the error tails, and the dependence on the downstream localizer.

\subsection{Scenario and parameter instantiation}
Spatially consistent channels are generated using QuaDRiGa \cite{jaeckel2014quadriga} under the 3GPP TR~38.901 indoor-NLOS model \cite{3gpp38901}. The carrier frequency is $3.5$~GHz. The BS is at $(0,0,3)$~m, the UE height is $1.5$~m, and the region $x\in[2,8]$~m, $y\in[-3,3]$~m is divided into 36 cells of size $1\times1$~m. Each cell contains an $8\times8$ lattice with $0.125$-m spacing, and the 2304 distinct coordinates are split without overlap into 1152 training, 576 validation, and 576 test coordinates. The experiment therefore evaluates interpolation at unseen coordinates within one spatially consistent radio map; it does not claim cross-environment generalization. For this controlled simulation, the exact earliest generated arrival is used as an idealized common relative-delay origin in both offline and online processing. This alignment removes a common timing offset but is not used as a ranging observable; robustness to offline/online timing-reference perturbations is left for future work.

The sounding waveform uses $M=64$ uniformly spaced pilots with $\Delta f_{\rm p}=3.120$~MHz and a first-to-last pilot span of $196.560$~MHz. Equations~\eqref{eq:ambiguity_period} and \eqref{eq:native_bin} therefore give $T_{\rm amb}=320.512821$~ns and $\Delta\tau_{\rm N}=5.008013$~ns; the twofold and threefold grids have spacings of $2.504006$ and $1.669338$~ns. This is a custom wideband OFDM sounding abstraction rather than a bit-exact standardized positioning-signal allocation. Every representation uses the full feature interval $[0,T_{\rm amb})$. No physical path is removed before pilot sampling, and no reconstructed delay bin in this interval is excluded from the feature support. Physical delays outside one ambiguity period contribute through the folding described in Section~II-B, and the same folded fingerprint is learned offline and used online. After reconstruction, the common shape normalization clips powers more than $B_{\rm dB}=50$~dB below the profile maximum to the zero feature level.

Two independent 20-dB noise realizations per training coordinate are retained as separate fingerprints, yielding 2304 training samples; two independent 20-dB realizations per validation coordinate are used only for model selection. Testing covers $\{5,10,15,20,25,30\}$~dB with 20 independent noise realizations per coordinate and SNR. Each plotted point averages the metric over all 576 unseen test coordinates and the 20 independent noise realizations, so the curves summarize receiver-noise variability within the same spatial environment. Noise is scaled per channel realization, and every front end uses paired channels and standardized noise draws. To isolate reconstruction and grid-mismatch effects, the generated thermal-noise variance is divided by the same $s_y$ used to normalize the observation and supplied only through $\bar\sigma_w^2$ as the lower bound in \eqref{eq:damped_updates}; the total effective residual variance remains EM-updated. We use $\rho=0.55$, at least 10 and at most 500 iterations, a relative stopping tolerance of $10^{-5}$, and pruning after iteration 15 at relative threshold $10^{-8}$ while retaining 4--64 active atoms. Top-1 cell accuracy and the Euclidean distance from the true coordinate to the predicted cell center are reported; for the adopted discrete test-coordinate set, perfect cell decisions have mean quantization error \QuantizationMeanErr~m.
\begin{table}[t]
\caption{Main simulation and implementation settings.}
\label{tab:settings}
\centering
\scriptsize
\setlength{\tabcolsep}{3.3pt}
\begin{tabular}{@{}p{0.34\columnwidth}p{0.59\columnwidth}@{}}
\toprule
Parameter & Setting\\
\midrule
Channel/map & QuaDRiGa indoor NLOS; $6\times6$~m\\
Train/validation/test & 1152/576/576 coordinates\\
Pilot grid & $M=64$, $\Delta f_{\rm p}=3.120$~MHz\\
First-to-last pilot span & 196.560~MHz\\
Ambiguity period & $T_{\rm amb}=320.512821$~ns\\
Native delay spacing & $\Delta\tau_{\rm N}=5.008013$~ns\\
Refined grid spacing & $q=2$: 2.504006~ns; $q=3$: 1.669338~ns\\
Feature support & full $[0,320.512821~\mathrm{ns})$\\
Feature clipping & post-reconstruction, below $-50$~dB\\
Noise repetitions & train/val.: $2/2$ at 20~dB; test: 20/SNR\\
EM-SBL & $\rho=0.55$; rel. $\gamma$ floor $10^{-10}$; $\sigma_0^2=10^{-4}$\\
Iteration/pruning & 10--500; tol. $10^{-5}$; prune after 15 at $10^{-8}$\\
\bottomrule
\end{tabular}
\end{table}
\FloatBarrier
\subsection{Front-end behavior and principal SNR results}
\begin{figure}[htpb]
\centering
\begin{minipage}[t]{0.49\columnwidth}
\centering
\includegraphics[width=\linewidth]{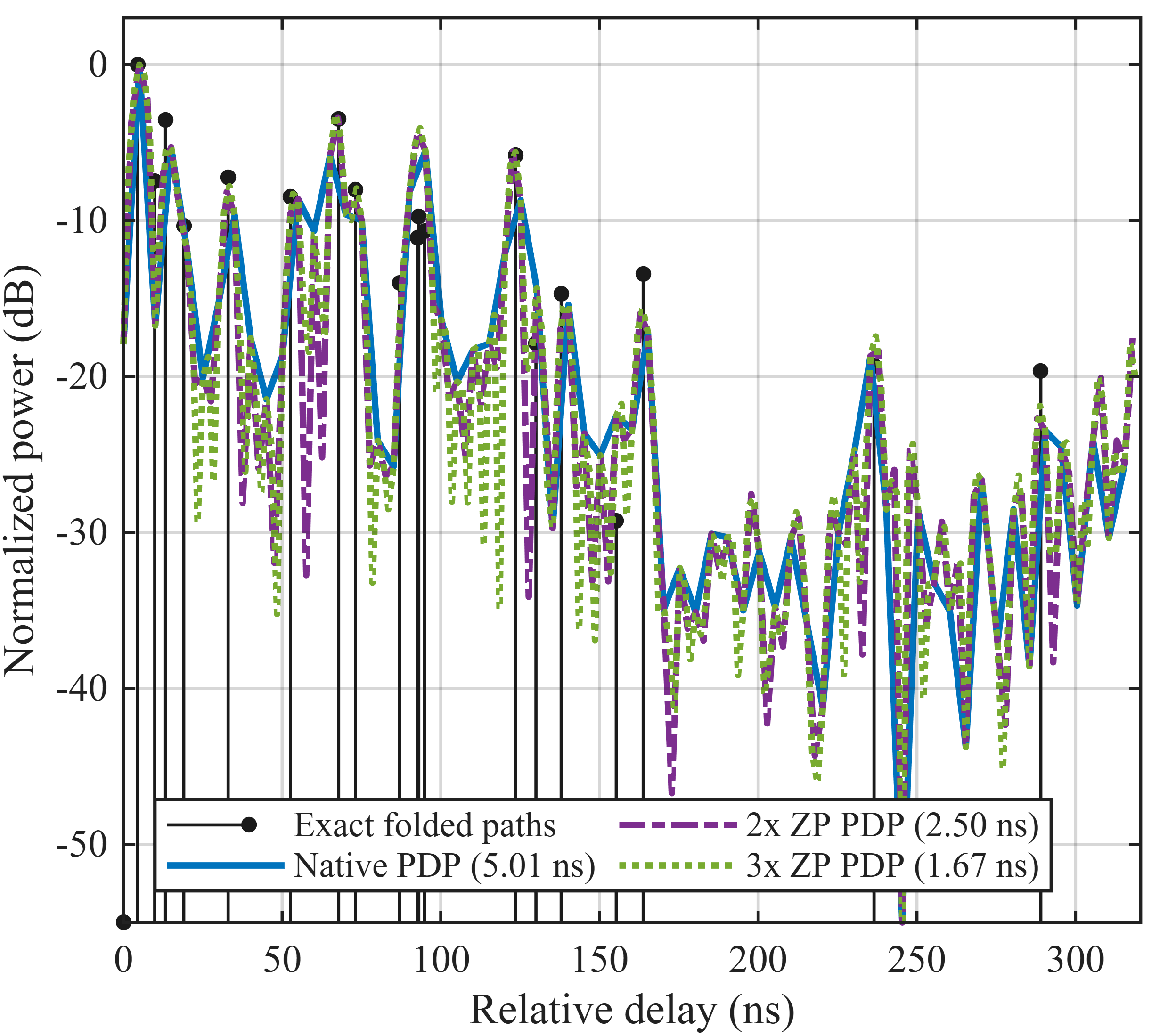}
{\scriptsize (a)}
\end{minipage}\hfill
\begin{minipage}[t]{0.49\columnwidth}
\centering
\includegraphics[width=\linewidth]{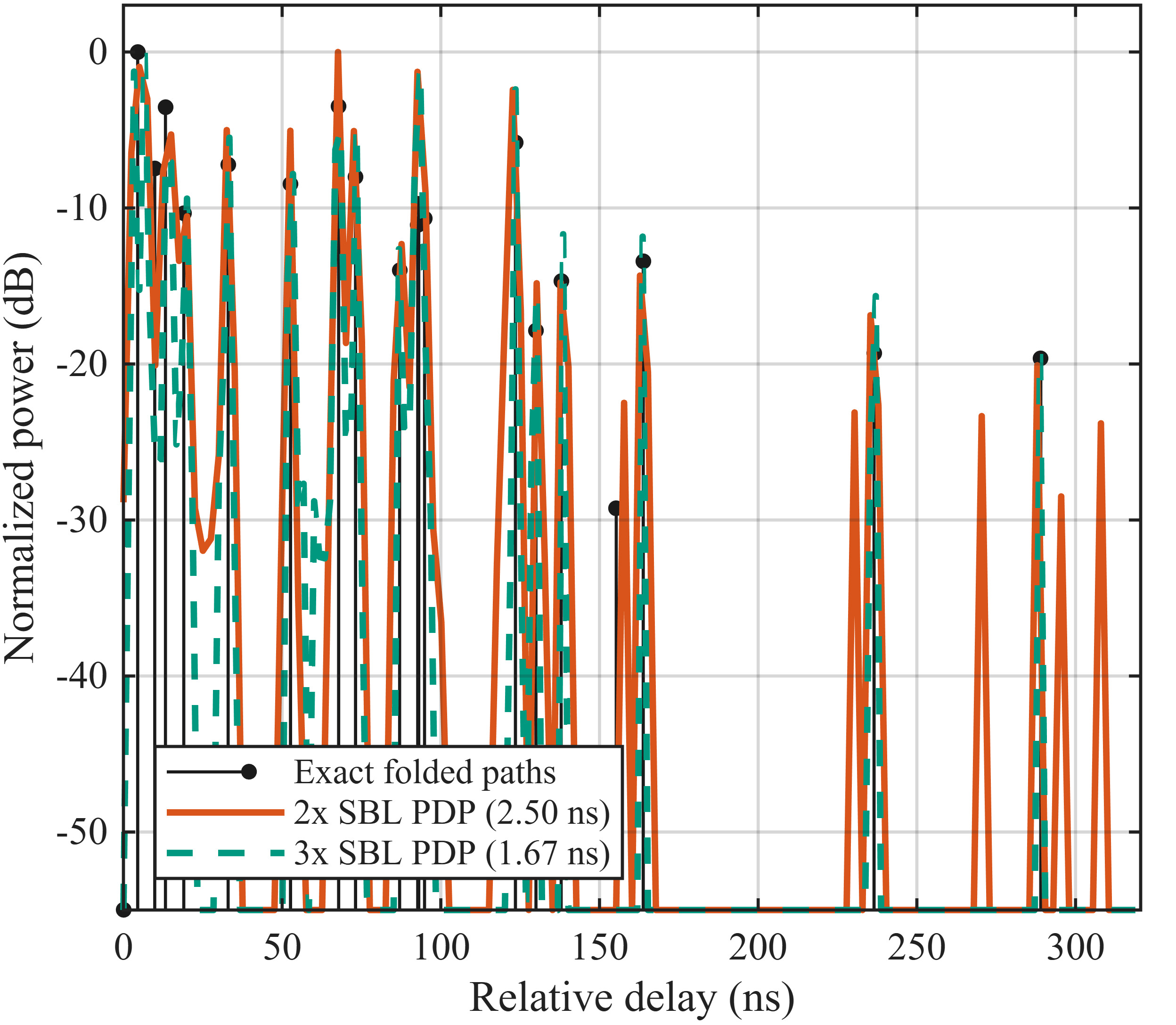}
{\scriptsize (b)}
\end{minipage}

\caption{Representative full-channel example at 20~dB. The black stems show the individual physical-path powers at their relative delays folded modulo $T_{\rm amb}$. (a) The native PDP is a sampled Dirichlet-blurred response, and twofold/threefold zero-padding only interpolate it. (b) Twofold and threefold SBL produce sharper posterior-power profiles; reconstructed atoms need not correspond one-to-one to physical paths.}
\label{fig:pdp_example}
\end{figure}

Fig.~\ref{fig:pdp_example} visualizes the signal-processing mechanism before presenting aggregate localization results. Nearby folded paths merge into broad Dirichlet lobes in the native PDP, while twofold and threefold zero-padding sample the same lobe and sidelobe pattern more densely. EM-SBL instead concentrates posterior power around a smaller set of delay atoms. Comparing SBL with its dimension-matched zero-padding baseline at both grid densities separates sparse refinement from interpolation.

The principal comparison fixes the localization rule to weighted kNN, generates the pilot observations from the complete physical channel, and retains all reconstructed bins over the ambiguity period. No path is removed before sampling and no delay bin is truncated afterward; only the common post-reconstruction 50-dB relative-power clipping maps weak reconstructed bins to zero. Pilot-grid aliasing is therefore not treated as an error to be removed: it is a deterministic part of the fingerprint and is reproduced identically during offline training and online testing.

\begin{figure}[!t]
\centering
\begin{minipage}[t]{0.49\columnwidth}
\centering
\includegraphics[width=\linewidth]{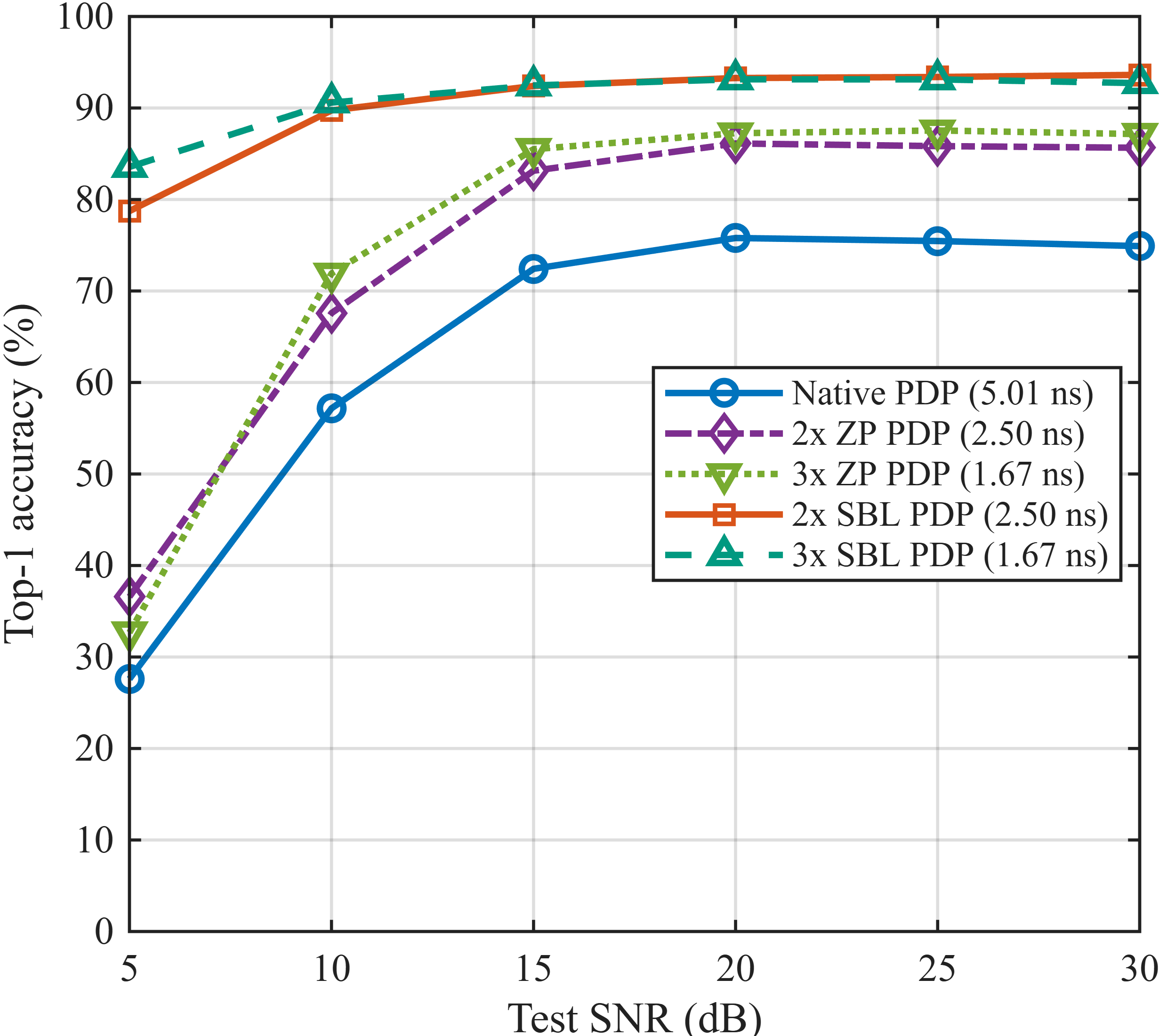}
{\scriptsize (a)}
\end{minipage}\hfill
\begin{minipage}[t]{0.49\columnwidth}
\centering
\includegraphics[width=\linewidth]{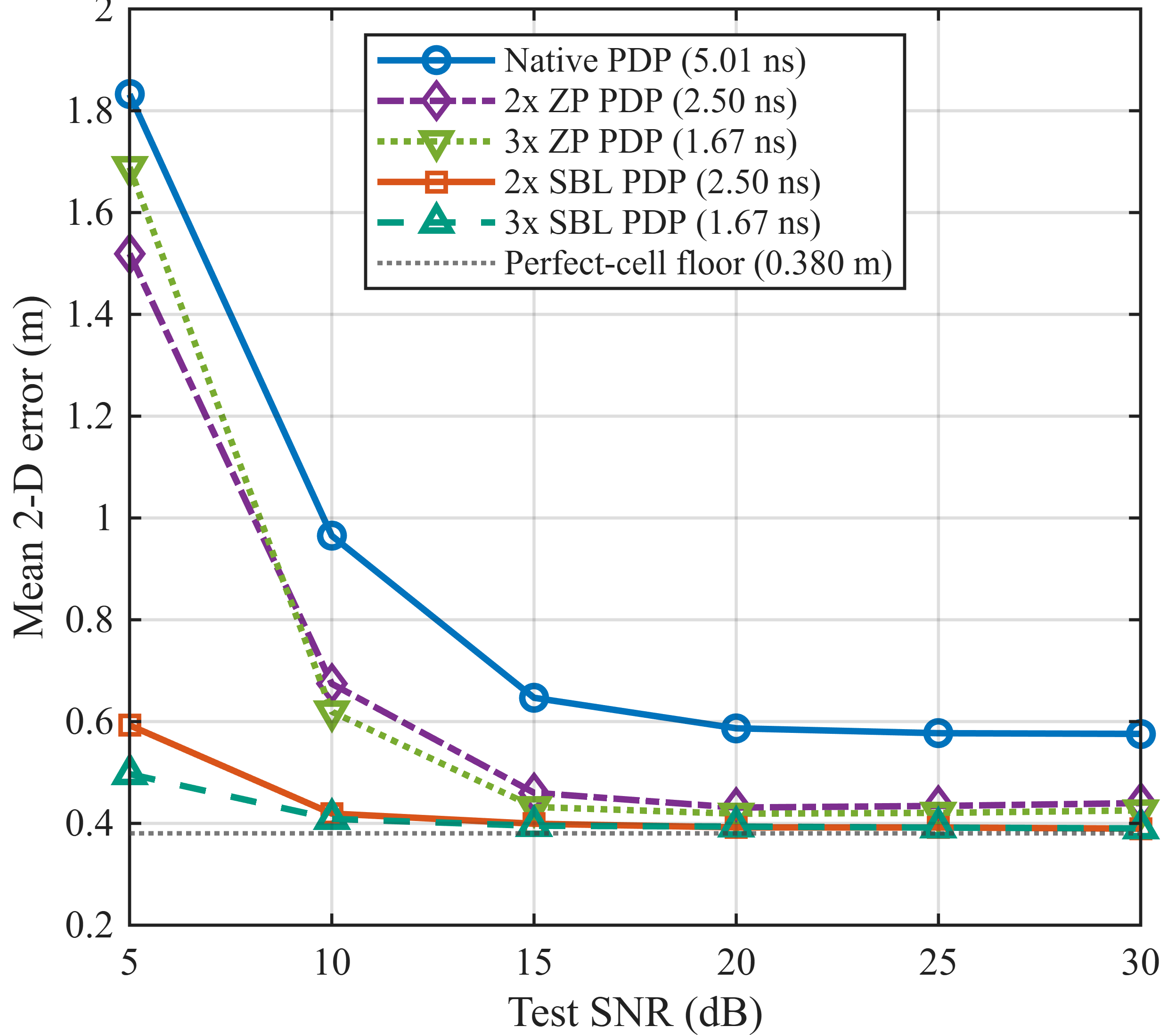}
{\scriptsize (b)}
\end{minipage}
\caption{Principal localization performance under the validation-selected weighted-kNN protocol. (a) Top-1 cell accuracy versus SNR. (b) Mean two-dimensional error versus SNR; the dotted reference is the perfect-cell quantization floor.}
\label{fig:snr_curves}
\end{figure}

Across all tested SNRs, each SBL grid outperforms its dimension-matched zero-padding baseline, whereas increasing the zero-padding factor alone gives only a modest improvement. At 20~dB, twofold SBL reaches \KNNSBLTwoAcc\% Top-1 accuracy, improving by 17.48 percentage points over the native PDP and by 6.03 points over threefold zero-padding; the corresponding mean errors are \KNNNativeErr, \KNNZPThreeErr, and \KNNSBLTwoErr~m. Threefold SBL is strongest at the lowest SNRs, while the two SBL grids are essentially equivalent from 15 to 30~dB. Twofold SBL therefore offers the preferable performance--dimension tradeoff.

\FloatBarrier
\subsection{Error tails and dependence on the localizer}
\begin{figure}[htpb]
\centering
\begin{minipage}[t]{0.49\columnwidth}
\centering
\includegraphics[width=\linewidth]{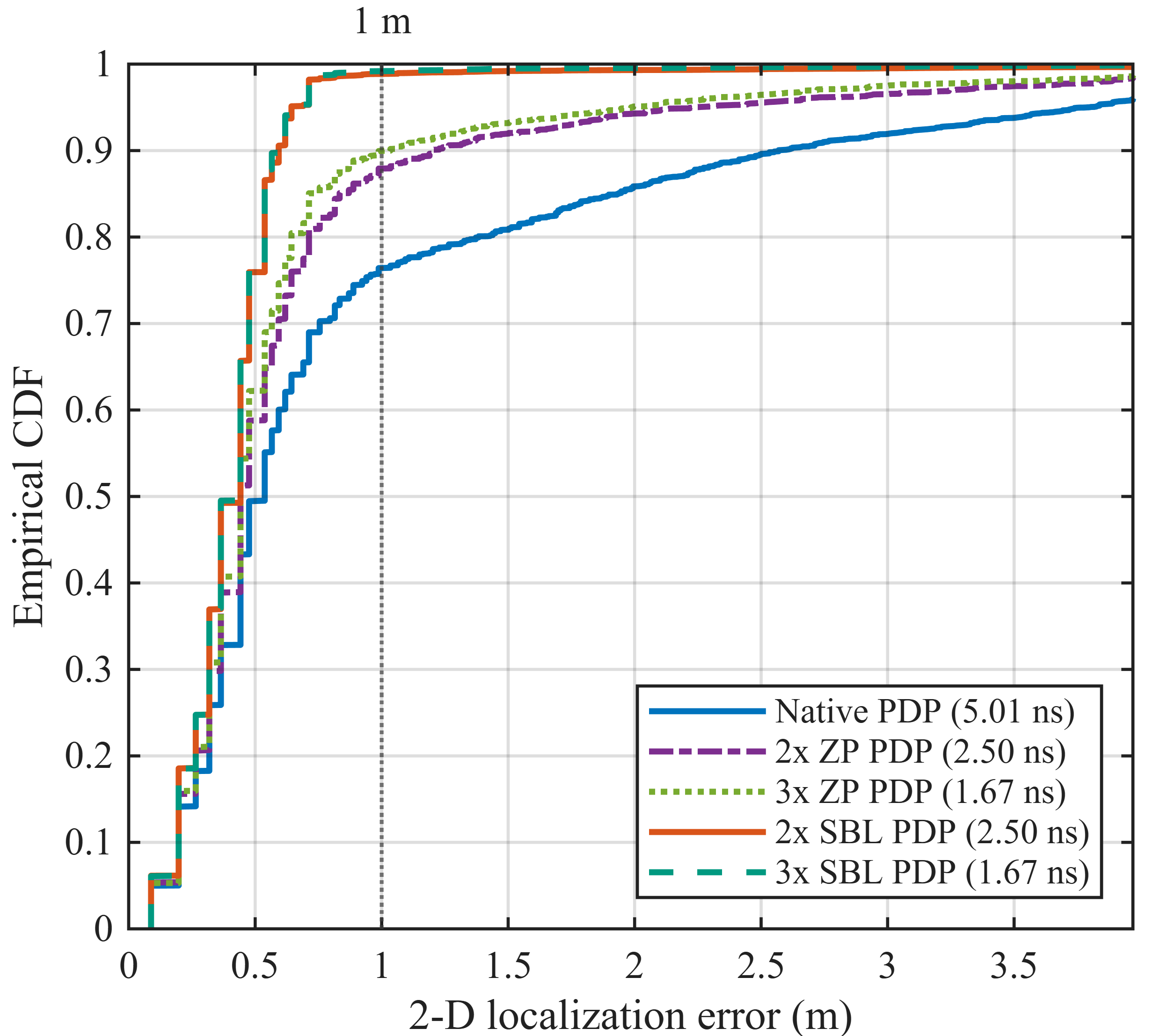}\\
{\scriptsize (a) 10 dB} 
\end{minipage}\hfill
\begin{minipage}[t]{0.49\columnwidth}
\centering
\includegraphics[width=\linewidth]{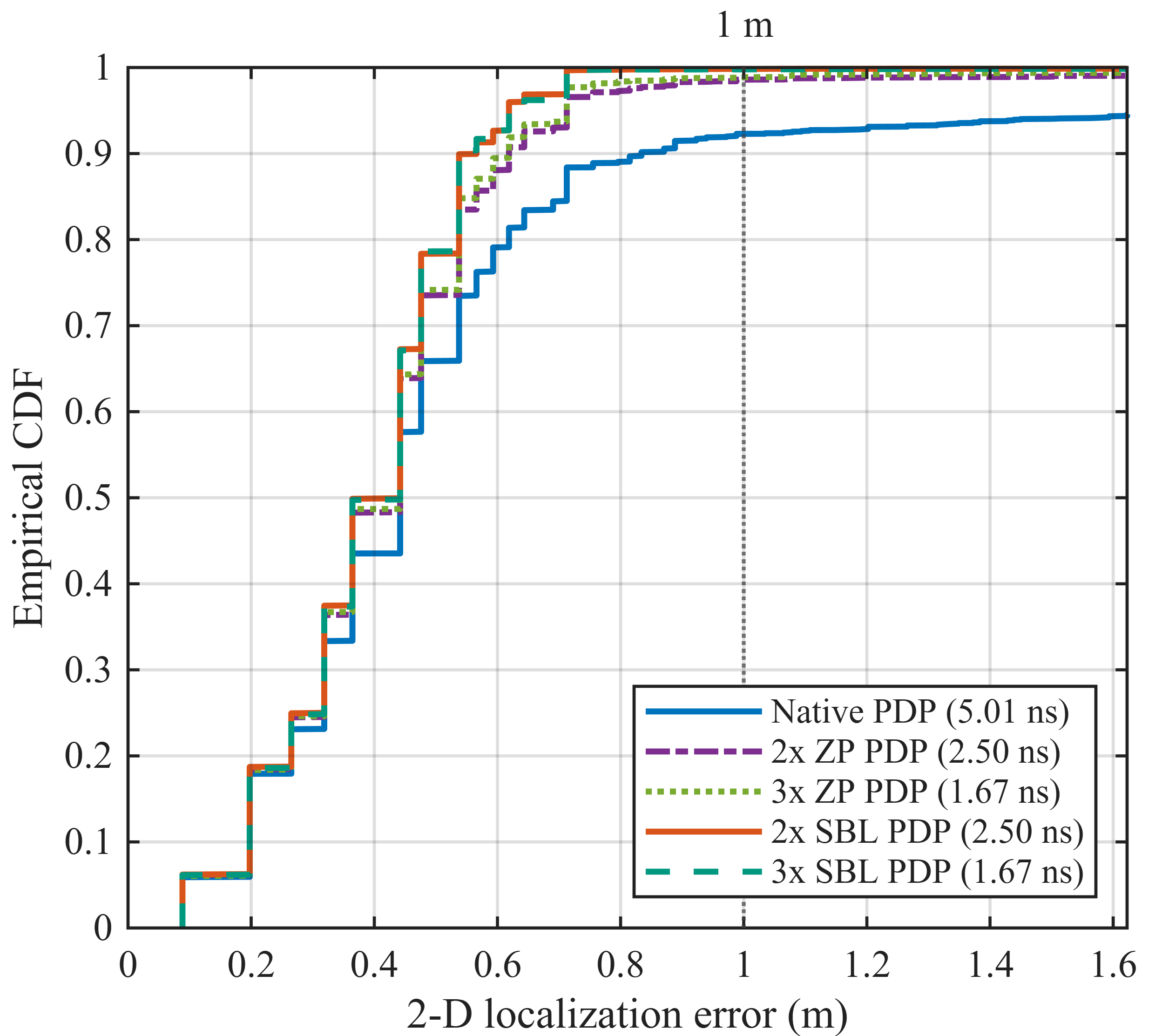}\\
{\scriptsize (b) 20 dB}
\end{minipage}
\caption{Empirical CDFs of the two-dimensional localization error under the validation-selected weighted-kNN protocol. The dashed vertical line marks 1~m; the panels use different horizontal ranges to preserve the visibility of their respective tails. The relative improvement from EM-SBL is visible both at 10~dB, where tail suppression is most important, and at 20~dB, where SBL approaches the cell-quantization floor.}
\label{fig:cdf_pair}
\end{figure}

The SNR curves in Fig.~\ref{fig:snr_curves} establish the principal gain. We next examine whether that gain also removes large localization failures. Fig.~\ref{fig:cdf_pair} compares the empirical error CDFs at 10 and 20~dB over ranges chosen to expose their respective tails. At 10~dB, both SBL curves place almost all errors below 1~m, while the native and zero-padded PDPs retain visibly heavier tails. The 20-dB CDF confirms the same ordering under a cleaner observation regime: SBL concentrates near the cell-quantization floor, threefold zero-padding slightly improves on twofold zero-padding, and the native PDP remains least concentrated. The gain therefore reflects tail suppression rather than only a small set of favorable queries.
\begin{table}[htpb!]
\caption{Cross-localizer results at 20~dB. Accuracy is in percent and error is mean two-dimensional distance in meters; boldface marks the best value at the reported precision for each localizer and metric.}
\label{tab:classifiers}
\centering
\scriptsize
\setlength{\tabcolsep}{1.5pt}
\begin{tabular}{@{}llccccc@{}}
\toprule
Localizer & Metric & Native & ZP2 & ZP3 & SBL2 & SBL3\\
\midrule
\multirow{2}{*}{Nearest centroid}
 & Acc. & \CentroidNativeAcc & \CentroidZPTwoAcc & \CentroidZPThreeAcc & \CentroidSBLTwoAcc & \textbf{\CentroidSBLThreeAcc}\\
 & Err. & \CentroidNativeErr & \CentroidZPTwoErr & \CentroidZPThreeErr & \CentroidSBLTwoErr & \textbf{\CentroidSBLThreeErr}\\
\cmidrule(lr){1-7}
\multirow{2}{*}{Weighted kNN}
 & Acc. & \KNNNativeAcc & \KNNZPTwoAcc & \KNNZPThreeAcc & \textbf{\KNNSBLTwoAcc} & \KNNSBLThreeAcc\\
 & Err. & \KNNNativeErr & \KNNZPTwoErr & \KNNZPThreeErr & \textbf{\KNNSBLTwoErr} & \KNNSBLThreeErr\\
\cmidrule(lr){1-7}
\multirow{2}{*}{Linear SVM}
 & Acc. & \LinearNativeAcc & \LinearZPTwoAcc & \LinearZPThreeAcc & \LinearSBLTwoAcc & \textbf{\LinearSBLThreeAcc}\\
 & Err. & \LinearNativeErr & \LinearZPTwoErr & \LinearZPThreeErr & \LinearSBLTwoErr & \textbf{\LinearSBLThreeErr}\\
\cmidrule(lr){1-7}
\multirow{2}{*}{RBF SVM}
 & Acc. & \RBFNativeAcc & \RBFZPTwoAcc & \RBFZPThreeAcc & \RBFSBLTwoAcc & \textbf{\RBFSBLThreeAcc}\\
 & Err. & \RBFNativeErr & \RBFZPTwoErr & \RBFZPThreeErr & \textbf{\RBFSBLTwoErr} & \textbf{\RBFSBLThreeErr}\\
\cmidrule(lr){1-7}
\multirow{2}{*}{Random forest}
 & Acc. & \RFNativeAcc & \RFZPTwoAcc & \RFZPThreeAcc & \RFSBLTwoAcc & \textbf{\RFSBLThreeAcc}\\
 & Err. & \RFNativeErr & \RFZPTwoErr & \RFZPThreeErr & \RFSBLTwoErr & \textbf{\RFSBLThreeErr}\\
\bottomrule
\end{tabular}
\end{table}

Table~\ref{tab:classifiers} tests whether the front-end conclusion depends on weighted kNN. Across nearest-centroid, linear-SVM, RBF-SVM, and random-forest inference, SBL consistently improves both accuracy and mean distance error over the native and zero-padded PDPs. Weighted kNN is retained in the principal curves because it achieves the best overall accuracy (93.27\%) and the lowest mean error (0.392~m), while keeping the localization stage transparent. The auxiliary linear and RBF SVMs use multiclass error-correcting output codes~\cite{cortes1995support,dietterich1995ecoc}, whereas the random forest uses bagged decision trees~\cite{breiman2001random}; one fixed implementation setting is shared across all representations, with no test-set tuning.
This cross-localizer consistency indicates that the dominant gain comes from the sampling-aware PDP representation rather than from a particular decision rule.
\FloatBarrier

\section{Conclusion}
This paper formulated relative-PDP fingerprint localization directly from finite OFDM pilot observations rather than assuming an ideal delay profile. The central premise is that indoor multipath carries environmental information: the PDP is used as a LoS-free radio-map fingerprint, not as a range or absolute-ToA measurement. Uniform pilot spacing produces a folded ambiguity-period representation, finite aperture produces Dirichlet blur, and continuous delays create grid mismatch. From identical pilots, the EM-SBL posterior second moment yielded substantially better cell accuracy and distance error than the native PDP and both dimension-matched zero-padding baselines. The complete ambiguity-period delay support was retained without window truncation, while a common post-reconstruction 50-dB relative-power clipping was applied to all front ends. SNR sweeps, the 10- and 20-dB error CDFs, and five localizers suggest that communication PDPs already contain sufficient environmental information for accurate indoor localization, provided that sampling-induced distortions are properly handled.

\FloatBarrier


\begin{thebibliography}{99}
\bibitem{bahl2000radar}
P. Bahl and V. N. Padmanabhan, ``RADAR: An in-building RF-based user location and tracking system,'' in \emph{Proc. IEEE INFOCOM}, Tel Aviv, Israel, Mar. 2000, vol. 2, pp. 775--784.

\bibitem{he2016wifi}
S. He and S.-H. G. Chan, ``Wi-Fi fingerprint-based indoor positioning: Recent advances and comparisons,'' \emph{IEEE Commun. Surveys Tuts.}, vol. 18, no. 1, pp. 466--490, 1st Quart., 2016.

\bibitem{kotaru2015spotfi}
M. Kotaru, K. Joshi, D. Bharadia, and S. Katti, ``SpotFi: Decimeter level localization using WiFi,'' in \emph{Proc. ACM SIGCOMM}, London, U.K., Aug. 2015, pp. 269--282.

\bibitem{vasisht2016chronos}
D. Vasisht, S. Kumar, and D. Katabi, ``Decimeter-level localization with a single WiFi access point,'' in \emph{Proc. 13th USENIX NSDI}, Santa Clara, CA, USA, Mar. 2016, pp. 165--178.

\bibitem{triki2006pdp}
M. Triki, D. T. M. Slock, V. Rigal, and P. Fran{\c{c}}ois, ``Mobile terminal positioning via power delay profile fingerprinting: Reproducible validation simulations,'' in \emph{Proc. IEEE VTC-Fall}, Montreal, QC, Canada, Sep. 2006, pp. 1--5.

\bibitem{triki2007nlos}
M. Triki and D. T. M. Slock, ``Mobile localization for NLOS propagation,'' in \emph{Proc. IEEE PIMRC}, Athens, Greece, Sep. 2007, pp. 1--4.

\bibitem{oktem2010pddp}
T. M. Oktem and D. T. M. Slock, ``Power delay Doppler profile fingerprinting for mobile localization in NLOS,'' in \emph{Proc. IEEE PIMRC}, Istanbul, Turkey, Sep. 2010, pp. 876--881.

\bibitem{ding2016pdp}
G. Ding, P. Chen, J. Tian, and Q. Zhao, ``Power delay profile based indoor fingerprinting localization system,'' in \emph{Proc. 18th Int. Conf. Advanced Commun. Technol.}, Pyeongchang, South Korea, Jan. 2016, pp. 324--329.

\bibitem{xiao2024ranging}
F. Xiao, Z. Zhao, and D. T. M. Slock, ``Multipath component power delay profile based ranging,'' \emph{IEEE J. Sel. Topics Signal Process.}, vol. 18, no. 5, pp. 950--963, Jul. 2024.

\bibitem{xiao2026pdpsensing}
F. Xiao, Z. Li, and D. T. M. Slock, ``Multipath component power delay profile based joint range and Doppler estimation for AFDM-ISAC systems,'' \emph{IEEE Trans. Commun.}, early access, Apr. 2026, doi: 10.1109/TCOMM.2026.3686728.

\bibitem{xie2019pdp}
Y. Xie, Z. Li, and M. Li, ``Precise power delay profiling with commodity Wi-Fi,'' \emph{IEEE Trans. Mobile Comput.}, vol. 18, no. 6, pp. 1342--1355, Jun. 2019.

\bibitem{bajwa2010ccs}
W. U. Bajwa, J. Haupt, A. M. Sayeed, and R. Nowak, ``Compressed channel sensing: A new approach to estimating sparse multipath channels,'' \emph{Proc. IEEE}, vol. 98, no. 6, pp. 1058--1076, Jun. 2010.

\bibitem{qiao2018sbl}
G. Qiao, Q. Song, L. Ma, S. Liu, Z. Sun, and S. Gan, ``Sparse Bayesian learning for channel estimation in time-varying underwater acoustic OFDM communication,'' \emph{IEEE Access}, vol. 6, pp. 56675--56684, 2018.

\bibitem{candes2014superresolution}
E. J. Cand{\`e}s and C. Fern{\'a}ndez-Granda, ``Towards a mathematical theory of super-resolution,'' \emph{Commun. Pure Appl. Math.}, vol. 67, no. 6, pp. 906--956, Jun. 2014.

\bibitem{tang2013offgrid}
G. Tang, B. N. Bhaskar, P. Shah, and B. Recht, ``Compressed sensing off the grid,'' \emph{IEEE Trans. Inf. Theory}, vol. 59, no. 11, pp. 7465--7490, Nov. 2013.

\bibitem{yang2013offgrid}
Z. Yang, L. Xie, and C. Zhang, ``Off-grid direction of arrival estimation using sparse Bayesian inference,'' \emph{IEEE Trans. Signal Process.}, vol. 61, no. 1, pp. 38--43, Jan. 2013.

\bibitem{uykan2026superresolution}
Z. Uykan, H. Al-Tous, O. Tirkkonen, and R. J{\"a}ntti, ``Fingerprinting localization based on super resolution CSI features,'' \emph{IEEE Access}, vol. 14, pp. 49750--49767, 2026.

\bibitem{schmidl1997sync}
T. M. Schmidl and D. C. Cox, ``Robust frequency and timing synchronization for OFDM,'' \emph{IEEE Trans. Commun.}, vol. 45, no. 12, pp. 1613--1621, Dec. 1997.

\bibitem{minn2003sync}
H. Minn, V. K. Bhargava, and K. B. Letaief, ``A robust timing and frequency synchronization for OFDM systems,'' \emph{IEEE Trans. Wireless Commun.}, vol. 2, no. 4, pp. 822--839, Jul. 2003.

\bibitem{tipping2001sbl}
M. E. Tipping, ``Sparse Bayesian learning and the relevance vector machine,'' \emph{J. Mach. Learn. Res.}, vol. 1, pp. 211--244, Jun. 2001.

\bibitem{wipf2004sbl}
D. P. Wipf and B. D. Rao, ``Sparse Bayesian learning for basis selection,'' \emph{IEEE Trans. Signal Process.}, vol. 52, no. 8, pp. 2153--2164, Aug. 2004.

\bibitem{jaeckel2014quadriga}
S. Jaeckel, L. Raschkowski, K. B{\"o}rner, and L. Thiele, ``QuaDRiGa: A 3-D multi-cell channel model with time evolution for enabling virtual field trials,'' \emph{IEEE Trans. Antennas Propag.}, vol. 62, no. 6, pp. 3242--3256, Jun. 2014.

\bibitem{3gpp38901}
3GPP, ``Study on channel model for frequencies from 0.5 to 100 GHz,'' 3GPP TR 38.901, ver. 19.2.0, Jan. 2026.

\bibitem{cortes1995support}
C. Cortes and V. Vapnik, ``Support-vector networks,''
\emph{Mach. Learn.}, vol. 20, no. 3, pp. 273--297, Sep. 1995.

\bibitem{dietterich1995ecoc}
T. G. Dietterich and G. Bakiri, ``Solving multiclass learning
problems via error-correcting output codes,''
\emph{J. Artif. Intell. Res.}, vol. 2, pp. 263--286, 1995.

\bibitem{breiman2001random}
L. Breiman, ``Random forests,'' \emph{Mach. Learn.},
vol. 45, no. 1, pp. 5--32, Oct. 2001.
\end{thebibliography}
\end{document}